\documentstyle[12pt,aps,epsf]{revtex}

\begin{document}




\title{
Gluelump Spectrum in the Bag Model
}

\vskip 1cm
\author{Gabriel Karl\footnote{permanent address: Dept of Physics, University of Guelph, Ontario, Canada} and Jack Paton}
\address{Dept of Physics (Theoretical), 1 Keble Rd. , Oxford, UK OX1 3NP}


\date{April 1999}
\maketitle

\vspace{0.5cm}
\begin{center}
{\bf Abstract}
\end{center}
\vspace {.2cm}

\begin{abstract}
We explore the ordering of the lowest levels in a simple bag model of the ``gluelump'' of Michael \cite{FM} and also discuss, again within the context of the bag model, the related problem of hybrid potentials in the limit of very small spacing between quark and anti-quark sources \cite{FM} \cite{Juge}.
\end{abstract}

\pacs{}
\newpage

\section{Introduction}
Foster and Michael \cite{FM} have studied numerically in lattice gauge models an interesting system consisting of a spinless static color octet source at the origin which modifies the gluonic field of pure color gauge theory in its vicinity. They call this system a ``gluelump'' and we follow their nomenclature. In \cite{FM} and several earlier publications referred to in \cite{FM} they have discussed the spectrum of this system. It is of interest to use this system as a testing ground for models of low energy QCD. Two models which have been much discussed in the literature are the bag model  and the flux tube model  and both may be applied to this system. We shall restrict ourselves here to the bag model. This would appear to be a particularly simple system to analyse in the bag model since the bag has a definite centre  unlike the case of light quarks or gluons confined in a bag. 
\\There are earlier discussions of the system in the bag model \cite{CS} \cite{Juge} with emphasis on the ground state configuration. We extend the earlier work to higher excited states and make a comparison with the lattice results. Physically the system would be of interest if heavy gluinos were to be produced experimentally and last long enough for them to have ther color neutralized. This was the motivation of the study by Chanowitz and Sharpe \cite{CS} who were interested in the ground state configuration of the system (which they called the ``glueballino''). Another physical connection is to the states of hybrid mesons in the limit of short distances between the quark and antiquark. Juge, Kuti and Morningstar \cite{Juge} have studied static hybrid potentials both on the lattice and in bag models and have shown that the bag gives a good representation of the lattice results at small quark-antiquark separation. The present discussion elaborates their work, as well as earlier work by Hasenfratz et al \cite{Has} and Ono \cite{Ono} at the special point $r=0$ and for very small values of $r$. We confirm in the bag model the results of Foster and Michael \cite{FM} for the quantum numbers of the low lying states, with plausible ordering  $ J^{PC}=1^{+-}$,$1^{--}$,  $2^{--}$,$3^{+-}$, $2^{+-}$.
\\The system we study has similarities to a glueball, which consists entirely of gluons. In the case of the {\it gluelump} one of the gluons is very heavy, and is located at the origin. The analogous heavy meson state, $Q{\bar{q}}$ has also been studied in the bag model \cite{Shuryak} and it was noted that the problem of centre of mass motion (and spurious excited states) disappears. 
\section{Bag Model Calculation}
In the bag model, the system consists (in SU(3) of color) of an octet color charge fixed at the origin and one or several cavity ``gluons'' neutralizing the color charge in a spherical cavity of radius $R$. The one gluon states have lowest energy and we shall study them first, but will later estimate roughly the energies of the lowest two-gluon states. With a single gluon the energy of the system will have three contributions: a volume term $4\pi R^3 \Lambda/3$ where $\Lambda$ is the bag constant, the energy of the gluon mode, $k$, and $V^C$, the Coulomb interaction energy between the gluon and the octet charge at the origin. Initially we neglect $V_C$ so that the gluon in the bag is described by a free field.
\\The dynamics of a massless vector field confined to a cavity is well known \cite{Close} and the boundary conditions for the color electric ${\bf E}$ and magnetic ${\bf B}$ fields are ${\bf E \cdot r} = 0$ and ${\bf B \times r = 0}$ at $r = R$. The fields ${\bf E}$ and ${\bf B}$ are solutions of the (vector) wave equation which may be represented as ${\bf L}Y_{lm}j_l$ with $Y_{lm}$ spherical harmonics, $j_l$ spherical Bessel functions and ${\bf L}= {\bf r \times \nabla}$.
There are two sets of solutions of the vector wave equation, one in which ${\bf E} \propto {\bf L}Y_{lm}j_l$ and  ${\bf B \propto \nabla \times} {\bf L}Y_{lm}j_l$,  called TE modes (because ${\bf E \cdot r} = 0$ automatically) and the other set, the TM modes, in which ${\bf E}$ and ${\bf B}$ are interchanged. From these forms one may derive the equations which determine $kR$ through the boundary conditions. For the TE modes $kR$ is a solution of the equation ${\frac{d}{dr}(rj_l(kr)) =0}$ at $r=R$, while for the TM modes $kR$ is a solution of $j_l(kr) = 0$ at $r = R$. The lowest roots of these equations are as follows:\\
TE1: $\:kR= 2.744,\, 6.117,... $ for $J^P=1^+$\\
TE2: $\:kR= 3.870,\, 7.443,... $ for $J^P=2^-$\\
TE3: $\:kR= 4.973,\, 8.772, ... $ for $J^P=3^+$\\
TM1: $\:kR= 4.493,\, 7.72\,,... $ for $J^P=1-$\\
TM2: $\:kR = 5.763,\, 9.09\,,... $ for $J^P=2^+$\\
Since they are all single gluon modes they have $C=-1$.\\
Therefore the five lowest modes, in increasing order of $kR$, are 
\[ J^P = 1^+ (kR=2.744),\: 2^- (kR =3.870),\: 1^- (kR=4.493),\: 3^+ (kR= 4.973),\: 2^+ (kR=5.763) \]
We note that the first three states ($1^+, 2^-, 1^-$) are the same as found on the lattice \cite{FM} but the order of the $2^-$ and $1^-$ states is reversed. However we have still to consider the color Coulomb interaction with the octet charge at the origin, and we shall see that this interaction may easily reverse the order of the $2^-$ and $1^-$ states, to bring agreement with the lattice results. We treat the color Coulomb interaction by perturbation theory. This is justified, for our parameter choice, by the small ratio of the energy shift to the spacing between levels of the same quantum numbers. (\cite{Shuryak})\\
The color Coulomb interaction between the central charge and the gluon field is a consequence of the non-Abelian nature of gluons. It is easy to compute it incorrectly! For example, if we were to take the volume integral of ${\bf E_1 \cdot E_2}$ where ${\bf E_1}$ is the field of the point color charge and ${\bf E_2}$ is
that of the gluon, we would get zero since since ${\bf E_1}$ is longitudinal while ${\bf E_2}$ is transverse, so that ${\bf E_1 \cdot E_2}$ vanishes. We first obtain the color charge density of the gluon field which is proportional to ${\bf E_2 \cdot E_2}$, average over the polar angles $\theta, \phi$ and integrate with $(Q/r)$ where $Q$ is the central charge. An earlier calculation of this interaction is reported by Ono \cite{Ono} who uses for the gluon the approximation of uniform color charge density inside the spherical cavity. In Ono's approximation all states shift by the same amount, so that we must go beyond this approximation to find the relative shift of levels.\\
To compute the (color) Coulomb interaction between the central octet charge and the gluon we use a  ``confined'' Coulomb potential which obeys the boundary condition that the potential vanishes at the boundary of the bag, $V(R)=0$; so the potential of a central Abelian charge $q$ is  $\frac{q}{4\pi}(r^{-1}-R^{-1})$. There is an additional factor of $-3$ associated with the color $SU(3)$. For consistency we also include the change in self-energy, due to the boundary conditions, of the central charge which also depends on $R$, and the self-energy of the gluon which is estimated in the approximation of averaging over angles. The constant term in the potential does not contribute to the final answer.
We shall illustrate the computation for the case of the lowest TE1 mode in which ${\bf E_{10}} = c{\bf L}Y_{10} j_1(kr)$ where $c$ is a normalization constant and we have chosen $(l,m) = (1,0)$ to simplify the calculation. We have immediately that $[{\bf E_{10}}]_z =0$ and, from angular momentum algebra,
\[ [{\bf E_{10}}]_x =-ic\frac{y}{r}\sqrt{\frac{3}{4\pi}}j_1(kr),\:\:
[{\bf E_{10}}]_y =+ic\frac{x}{r}\sqrt{\frac{3}{4\pi}}j_1(kr). \]
Therefore ${\bf E_{10}^2} = c^2(1-\cos^2 \theta)\frac{3}{4\pi} j_1^2(kr)$ and integrating over all angles one obtains $<{\bf E_1}^2>$ proportional simply to $j_1(kr)^2$,  leading to the following expression for the color Coulomb interaction energy:
\[ V^C = -3 \alpha_s \frac{kR}{R} \frac{\int_0^{kR}x^2 (1/x - 1/(kR)) dx j_1(x)^2}{\int_0^{kR}x^2 dx j_1(x)^2} \]
where the factor $-3 = <{\bf F_1 \cdot F_2>_1}$ is the coupling of the color octets into a singlet and $\alpha_s$ is the color fine structure constant. The factor $kR$ in this case is 2.744, but the formula also applies to TE2, TE3 modes where we replace $j_1$ by $j_2$ or $j_3$ and use the apropriate value of $kR$. Incorporating also the self-energies detailed above, this leads to the following estimate for the first three TE modes:
\[ V^C(TE1) = -2.36 (-1.133) \alpha_s/R, \] \[  V^C(TE2) = -2.13 (-0.837) \alpha_s/R,\] \[V^C(TE3) = -2.01 (-0.680) \alpha_s/R \]
where the bracketed numbers correspond to neglecting the self energies. These coefficients were obtained by numerical integration of the spherical Bessel functions in the expression above. They diminish with the angular momentum $l$ as the color charge density shifts further from the origin with increasing $l$.\\
For the TM modes, where the color electric field is proportional to ${\bf \nabla \times} {\bf L }Y_{lm}j_l$, the computation is a little more tedious, and we find for the TM1 mode
\[ <{\bf E_1}^2> = c^2\left(\frac{3j_1^2}{r^2} + {{j_1^\prime}}^2 + \frac{2j_1\prime{j_1}}{r}\right)  \] 
and a similar expression with different coefficents for the TM2 mode, from which we compute, including self-energies (excluding self-energies in brackets):
\[ V_C(TM1) = -4.74 (-4.04)\frac{\alpha_s}{R},\:\: V_c(TM2) = -3.32(-2.40)\frac{\alpha_s}{R}. \]
The Coulomb interactions are larger than for the TE modes because the charge density in the case of the TM modes is closer to the origin.
\\We now have the ingredients to compute the energy of the bag by finding the extremal value of R. In this way one obtains
\[ E = \frac{4}{3} (4\pi \Lambda)^{\frac{1}{4}} (\alpha_{nj} - \kappa \alpha_s)^{\frac{3}{4}} = .79 (\alpha_{nj} - \kappa \alpha_{s})^{\frac{3}{4}} {\rm GeV}  \]
where $\Lambda$ is the bag constant, $\alpha_{nj} =(kR)_{nj}$, $\kappa$ is the coefficent of $\alpha_s$ in the color Coulomb interaction. (For example, for the TE1 mode, $\alpha_{nj}=2.744$ and $\kappa =+2.36$). We have followed Juge et al \cite{Juge} and taken ${\Lambda}^\frac{1}{4} =.315 {\rm GeV}$, $\alpha_s = .23$, and so obtain:
\[ E(1+) = 1.43 {\rm GeV}, E(2-) = 1.97 {\rm GeV}, E(3+) = 2.44 {\rm GeV}\] 
\[ E(1-) = 1.98 {\rm GeV}, E(2+) = 2.64 {\rm GeV},\] 
We note that the states of $J^P = 2^-, 1^-$ are now essentially degenerate for this value of $\alpha_s$.  For any larger values of $\alpha_s$ their order is reversed and the order of the first three states is now as found on the lattice \cite{FM}.\\
Before working out further one gluon states one should consider the position of two gluon states, which can also neutralize the central octet charge. The energy of a two gluon state has contributions from the energies of each of the gluons, their color electric interaction with the central charge, their mutual color electric interaction, and a color magnetic interaction between them. There are also the source and gluon self-energies as in the case of the one gluon states. The lowest such state is composed of two TE1 gluons, since TE1 is the lowest single gluon state. The only change which occurs relative to the case of the one gluon state comes from the fact that the interaction between two octet states (either gluon with central charge or between the two gluons, has a different color factor $<{\bf F_1 \cdot F_2}>_{\bf 8} = -3/2$, half the size of the factor $<{\bf F_1 \cdot F_2}>_{\bf 1} = -3$, since the  two octets are combined in an octet rather than a singlet. The interaction of each gluon with the central charge is the same as in the single gluon case apart from this factor of $\frac{1}{2}$ but there are two gluons. The interaction energy between the two gluons is a little cumbersome to estimate, so we shall here make the approximation of assuming a uniform color charge distribution for each of the gluons for this evaluation. In this approximation the color Coulomb interaction of the two gluons with the central charge is just the same as that of the single gluon previously ($-1.13 \alpha_s/R$) and the interaction between the two gluons is 
\[ \frac{1}{5}\frac{\alpha_s}{R}\left(-\frac{3}{2}\right)= -.3 \,\alpha_s/R
\] 
Combining these results and the relevant self-energies, we estimate the two gluon states to have an energy of $.79 \,{\rm GeV}\: (5.5-2.4\alpha_s)^\frac{3}{4}$ $\simeq2.6 \,{\rm GeV}$. This estimate neglects the color magnetic interaction which will split the the various $J^P$ states from each other. With two $J^P=1+$ states one expects $0^+$, $1^+$ and $2^+$ states. Presumably the $0+$ state lies lowest with the $1+$ and $2+$ higher as with glueballs, though the glueball spectrum in the bag model does not correlate too well with that determined on the lattice. Therefore, one finds that the lowest states of the gluelump spectrum are $1^{+-}, 1^{--}, 2^{--},  3^{+-}, 0^{+ +},2^{+-}$ with the relative positions of the $1^{--}$ and $2^{--}$ states and of the $0^{++}$  and $2^{+-}$ state not well determined by our estimates because of the uncertainty in $\alpha_s$.\\
\section{Hybrid Potentials At Small Quark-Antiquark Separation}
We next consider the effect of replacing the central octet charge by a pair of static color triplet and anti-triplet sources (in overall octet state) with the distance between the two sources being very small compared to the bag radius $R$. This system is equivalent to the addition of a small color octet dipole moment on top of the color octet charge. It is easy to see that the $J=1$ states will split into a ``molecular" $\Sigma$ state (coming form the $m=0$ component) and a molecular $\Pi$ doublet (coming from the $m=\pm1$ components) where $m$ is defined relative to a $z$-axis along the direction of the dipole.  We argue below that the energy of these ``molecular" states varies quadratically as a function of the distance  $\cal R$ between the triplet and anti-triplet with different coefficents for $\Sigma$ and $\Pi$ making the $\Pi$ doublet lower. It should be noted that we are neglecting at this point the color electrostatic interaction between the the triplet anti-triplet pair (which goes like $\alpha_s/{6\cal R}$) but this is the same in both $\Pi$ and $\Sigma$.
\\The action of the the dipole operator $\mu_z^{el}$ on the gluon wave function is equivalent to multiplication by the $z$-coordinate, where the $z$-axis is the axis of the dipole. This changes the multipolarity of the the gluon state, and in particular the parity of the gluon state flips. It is clear that all diagonal matrix elements vanish because of parity. Therefore the first non-zero contributions to the energy are in second order in the dipole operator, and therefore proportional to ${\cal R}^2$. We can evaluate approximately the coefficent of this quadratic term by using a single average energy denominator $\Delta E$, which we remove from the sum and use closure. In this way we obtain:
\[ E({\cal R}) = E(0) + \frac{9}{4} \alpha_s^2 <\frac{\cos^2 \theta}{r^4}>\frac{1}{\Delta E} {\cal R}^2   \]
where the expectation value is over the lowest state TE1,$m$ in the case of the ground state. One obtains two different values for $m=0$ and $m=\pm 1$. For the angular integrals we obtain  $(2/5)$ for the $\Pi$ states ($m=\pm1$) and $1/5$ for the $\Sigma$ state ($m=0$). Recalling that for the ground state the energy denominator $\Delta E$ is negative, we find that the lower molecular state is the $\Pi$ state and the next excitation is the $\Sigma$, in agreement with the lattice computations of \cite{Juge}. Evaluating numerically the integrals over spherical Bessel functions we estimate the splitting between the $\Sigma$ and $\Pi$ states to be about $0.2\left(\frac{\cal R}{R_{\rm bag}}\right)^2$ in GeV, which is similar to the lattice estimate at short distances. \cite{Morning}
\section{Concluding Remarks}
Our bag model results for the order of levels in the gluelump spectrum are gratifyingly close to those of the lattice. The actual magnitude of the splitting between the lowest two levels is given as $.35$ GeV in \cite{Michael2}. With the bag parameters we have chosen it is somewhat larger, $.54$ GeV. It is also amusing that our crude bag model estimate of the $\Sigma-\Pi$ splitting of the lowest hybrid potentials also agree qualitatively with the lattice results. \cite{Michael3}\\
Some lattice gauge results for the gluelump spectrum also exist in the case of $SU(2)$ of color \cite{Michael3}. The corresponding bag model calculation differs from the above only by the change in group theory factors. Normalizing to the string tension which in the bag model is given by 
\[ SU(3): \sigma = \sqrt{\frac{32 \pi \alpha_s \Lambda}{3}} \],
\[ SU(2): \sigma = \sqrt{6 \pi \alpha_s \Lambda} \]
 and noting that  the ``color Coulomb'' potential has about the same strength in SU(2) as in SU(3), we have \[ \Lambda(SU(2))=\Lambda(SU(3)),\:\:\frac{3}{4}\alpha(SU(2))\simeq\frac{4}{3}\alpha(SU(3))  \]
so that there is no appreciable change in the bag model spectrum in going from  $SU(3)$ to $SU(2)$ (apart from the fact that the two gluon state must now have $C=-$ only). There is a suggestion from the lattice \cite{Michael2} that the separation in $SU(2)$ may be less than in $SU(3)$.\\
\acknowledgements{We thank PPARC for financial support and GK also thanks NSERC, Canada for support.}

\end{document}